\documentclass{calor2002}

\usepackage{graphicx}   
\usepackage{color}      
\DeclareGraphicsExtensions{.eps,.ps}
\usepackage{layout}
\usepackage[dvips]{epsfig}
\usepackage{amssymb,xspace}
\usepackage{psfrag}

\begin{document}

\title{The Jet Calibration in the H1 Liquid Argon Calorimeter}

\author{C.~Schwanenberger}

\address{Deutsches Elektronen-Synchrotron DESY, Notkestra{\ss}e 85,\\ 
D-22607 Hamburg, Germany\\ 
E-mail: schwanen@mail.desy.de}

\author{on behalf of the H1 Collaboration}


\maketitle

\abstracts{The jet calibration of the Liquid-Argon-Calorimeter of the H1
 Detector at HERA is described. In the measurement of high jet transverse
 energies systematic uncertainties as low as $2\%$ can be
 reached in deep inelastic scattering with a high photon virtuality
 ($Q^2$) and in photoproduction. Furthermore,
the concept of a new
energy weighting scheme of H1 is
 presented. First applications with a high $Q^2$ neutral current deep
 inelastic scattering
 sample show that the resolution of the balance in
 transverse momentum between the hadronic system and the electron is
 improved.}


\section{Introduction}

In the H1 Experiment\cite{bib:h1} at HERA a Liquid Argon (LAr) Calorimeter
is used to measure the energies of particles emerging from the interaction
of 27.5 GeV positron (or electron) and 920 GeV (or 820 GeV) proton beams
over an angular range of $4^o \le \theta \le 154^o$.\footnote{H1 uses a
  right-handed coordinate system, where the direction of the proton beam
  defines the positive $z$-direction. The polar angle $\theta$ is measured
  with respect to this direction.}

The H1 Liquid Argon Calorimeter\cite{bib:lar} is a non-compensating
sampling calorimeter which is divided into 8 self supporting wheels each
built out of 8 octants in the barrel part or two half shells in the forward
part. It consists of electromagnetic and outer hadronic sections. In the
electromagnetic modules lead is used as absorber material which adds up to
20 -- 30 radiation lengths ($X_0$). The hadronic part is built out of
stainless steel absorber plates which corresponds to
a total of 
4.5 -- 8 interaction
lengths ($\lambda$) including the electromagnetic section.


\section{Calorimeter Calibration using H1 Physics Data}

Knowledge of the LAr calorimeter energy scale can be improved using neutral
current (NC) deep inelastic scattering (DIS)
physics data in off-line analysis. With the increased amount of data
collected in the past years the precision in this calibration method
has reached the design level.

\subsection{Energy calibration with DIS data}

The over-constrained kinematics of NC DIS events at HERA
allow the prediction of the energy $E_e$ of the scattered
electron\footnote{If not particularly emphasized, {\it electron} can mean
  either an electron or a positron.} from the electron beam energy
${E_{e-beam}}$, the
scattering angle of the electron $\theta_e$ and the effective angle of the final state
$\theta$ with the double angle method (DA)\cite{bib:da}:

\begin{equation}
E_e(\theta_e, \theta) = \frac{2 \cdot {E_{e-beam}}\cdot
   \sin{\theta}}{\sin{\theta_e}+\sin{\theta}-\sin({\theta_e}+{\theta})}\;\; .
\end{equation}

The remaining final state can be a hadronic shower (DIS-DA) or a
photon (QED-Compton-DA).
Using the predicted energies $E_e$, position dependent calibration
factors for the electromagnetic scale are derived.


The hadronic energy scale can be adjusted using
the known electron energy.
The scale correction factors
for the electromagnetic and hadronic sections
 are obtained wheelwise from the ratio
of transverse momenta of 
the calibrated
electron and the hadronic final state.

\begin{figure}[th]
\centerline{\epsfig{file=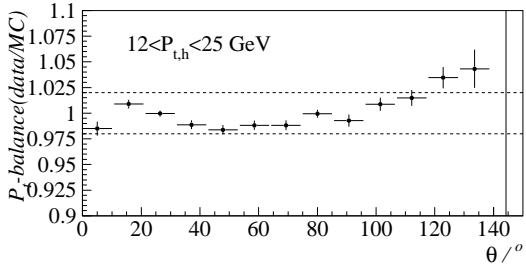,width=0.6\textwidth,bbllx=30,bblly=655,bburx=540,bbury=400,clip=}}\vspace{3.5cm}
\caption{The ratio of data and Monte Carlo
  prediction for the $p_t$-balance ratio as a function of the mean hadronic scattering
  angle $\theta$, for NC DIS with high momentum
  transfer $Q^2$ and for hadronic transverse momenta between 12 and 25
  GeV. \label{fig:ptbal}}
\end{figure}

The uncertainty of the hadronic energy scale is defined by the difference
between the correction factors in data and Monte Carlo simulation. As can be
seen in Fig.~\ref{fig:ptbal}, 
the uncertainties are below $2\%$ in a wide angular range of calorimeter acceptance. Larger
deviations can be found in the backward part where mostly low energy
hadrons are accepted
and where energy leakage occurs due to the missing hadronic section in the most backward wheel.

\subsection{Dijet Data in Photoproduction}

\begin{figure}[th]
   \begin{picture}(-10,220)
     \psfrag{tity}[][][1][0]{$\frac{E_{t,\rm max}}{E_{t,\rm rest}}$}
     \psfrag{titx}[][][1][0]{\hspace{2cm}${E_{t,\rm max}}$ (GeV)}
     \put(30,100){\epsfxsize=4.in\epsfbox{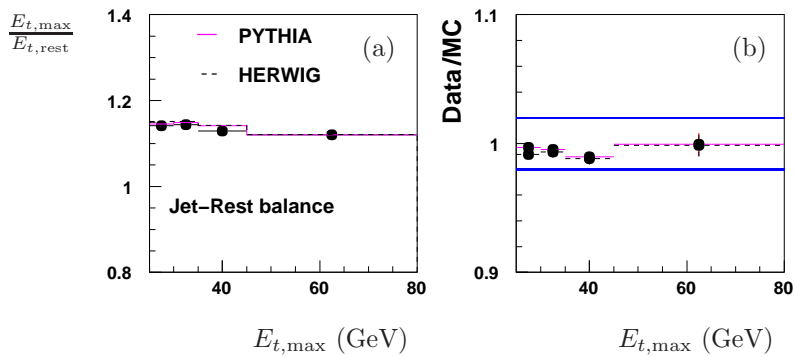}}
     \put(155,210){(a)}
     \put(295,210){(b)}
     \put(213.5,186.5){\color{blue}\line(1,0){101.1}}
     \put(213.5,167){\color{blue}\line(1,0){101.1}}
   \end{picture}
\vspace{-3.5cm}
\caption{(a) The ratio of the transverse energy of the highest energy jet
  and the transverse energy of the rest for data and two different Monte
  Carlo simulations as a function of the transverse energy of
  the highest energy jet. (b) The ratio of data and
  Monte Carlo prediction. \label{fig:etbal}}
\end{figure}

The calibration procedure obtained from DIS data is applied to the
measurement of the dijet cross section in
photoproduction.\cite{bib:dijet} The comparison of the ratio of
the transverse energy of the highest energy jet $E_{t,\rm max}$ and the transverse energy of the
rest\footnote{The {\it rest} is given by the energy of the second jet plus
remaining energies in
the event.} $E_{t,\rm rest}$ for data and Monte Carlo simulations
indicates the quality of the hadronic calibration. This ratio is shown in 
Fig.~\ref{fig:etbal} (a) as a function of the highest energy jet.
The  ratio of $E_{t,\rm max}/E_{t,\rm rest}$ in data 
and simulations, which is shown in
Fig.~\ref{fig:etbal} (b),  is consistent with uncertainties of the
hadronic energy scale below $2\%$.\footnote{This ratio is larger 1 due to
  the bias from the selection of the jet of highest $E_t$ and losses for
  $E_{t,\rm rest}$ in the beam pipe.} Detailed 
studies\cite{bib:sascha} demonstrate that at large transverse momentum these scale uncertainties are independent of
the angular distribution and the mass of the jets as well as of
different data selections such as direct, resolved or diffractive
processes.


\begin{figure}[th]
\vspace{-.2cm}
\centerline{\epsfxsize=3.2in\epsfbox{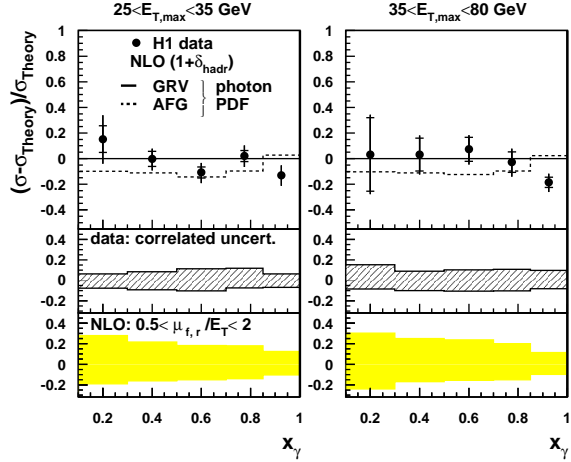}}   
\vspace{-.6cm}
\caption{The $x_{\gamma}$ dependence of the relative 
difference of the measured dijet cross sections ($Q^2< 1\,\mbox{GeV}^2$)
 from 
 the NLO prediction, with hadronization corrections applied
 (here $\sigma_{Theory}$).
  The symbol $\sigma$ stands for $d \sigma / d x_{\gamma}$.
  Figures a) and b) show the relative difference
  for a lower $E_{t,\rm max}$
  and a higher $E_{t,\rm max}$ region respectively.
  The inner
  error bars denote the statistical error, the outer error bars
  denote  
  all statistical and uncorrelated
  systematic errors of the data added in quadrature.
  The correlated systematic errors are shown in the middle plots
  as a shaded band.
  The bands in the lower plots
  show the renormalization and factorization
  scale uncertainties of this NLO prediction.\label{fig:dijet}}
\vspace{-.5cm}
\end{figure}

In order to illustrate the impact of these uncertainties on a cross section
measurement, 
the relative difference between the measured and the theoretical cross
sections\cite{bib:nlo} is given in Fig.~\ref{fig:dijet}, as a function of $x_{\gamma}$,
the longitudinal momentum fraction of the photon taken by the interacting
parton. The correlated errors due to the uncertainty in the calorimeter
energy scales are shown as a shaded band. Fig.~\ref{fig:dijet} shows that
the assumed scale uncertainties in the next-to-leading order QCD
calculation are the dominant source of uncertainties in the comparison of
data and theory.

\section{Towards a New Energy Weighting Scheme}

Because the H1 LAr calorimeter is non-compensating, a 
software weighting method\cite{bib:weighting} is applied for the energy
reconstruction.
In order to overcome some deficits in the low energy regime of the
current energy weighting scheme a new weighting procedure
was studied using CERN test beam data.\cite{bib:cigdem}

The reconstructed energy $E^i_{\rm rec}$ in a calorimeter cell
$i$ is derived from the measured cell energy $E^i_0$ by

\begin{equation}
E^i_{\rm rec} =\omega\; (\,
E^i_0/{\it Vol}^i , E_{\rm group}\,)\cdot E^i_0 \;\; .
\end{equation}

The weighting factor $\omega$ depends on the deposited energy density
$E^i_0/{\it Vol}^i$ ($Vol^i$ being the cell volume) which is different for an
electromagnetic and a hadronic deposition, and on the total energy to
be reconstructed (energy of the group of selected clusters $E_{\rm group}$). 
The latter accounts for the fact that both the relative
difference   
between the energy response to electrons and to pions\footnote{For instance, at an energy
of 10 GeV, the response to electrons is a factor 1.35 larger than the
response to pions.} and the fluctuation of the electromagnetic
component in the
hadronic shower\footnote{This is due to neutral pion production in nuclear
  reactions in the hadronic component of the shower.} depend on the
energy. 

The weighting factors $\omega$ are tabulated
and derived wheelwise,
separately for
the electromagnetic and the hadronic parts of the whole calorimeter, using
a more detailed simulation of single pions as standard for H1 physics
analyses.\cite{bib:joerg} Furthermore,
noise corrections are applied.
The method is
valid from the highest energies down to the noise level.
For the final calibration real DIS data
are used: the $p_t$-balance is adjusted wheelwise to $p_t^{\rm had} / p_t^e
\rightarrow 1$.

\begin{figure}[th]
   \begin{picture}(0,318)
     \psfrag{tity}[][][1][0]{$\frac{E_{t,\rm max}}{E_{t,\rm rest}}$}
     \psfrag{titx}[][][1][0]{\hspace{2cm}${E_{t,\rm rest}}$ (GeV)}
     \put(-30,100){\epsfxsize=2.5in\epsfbox{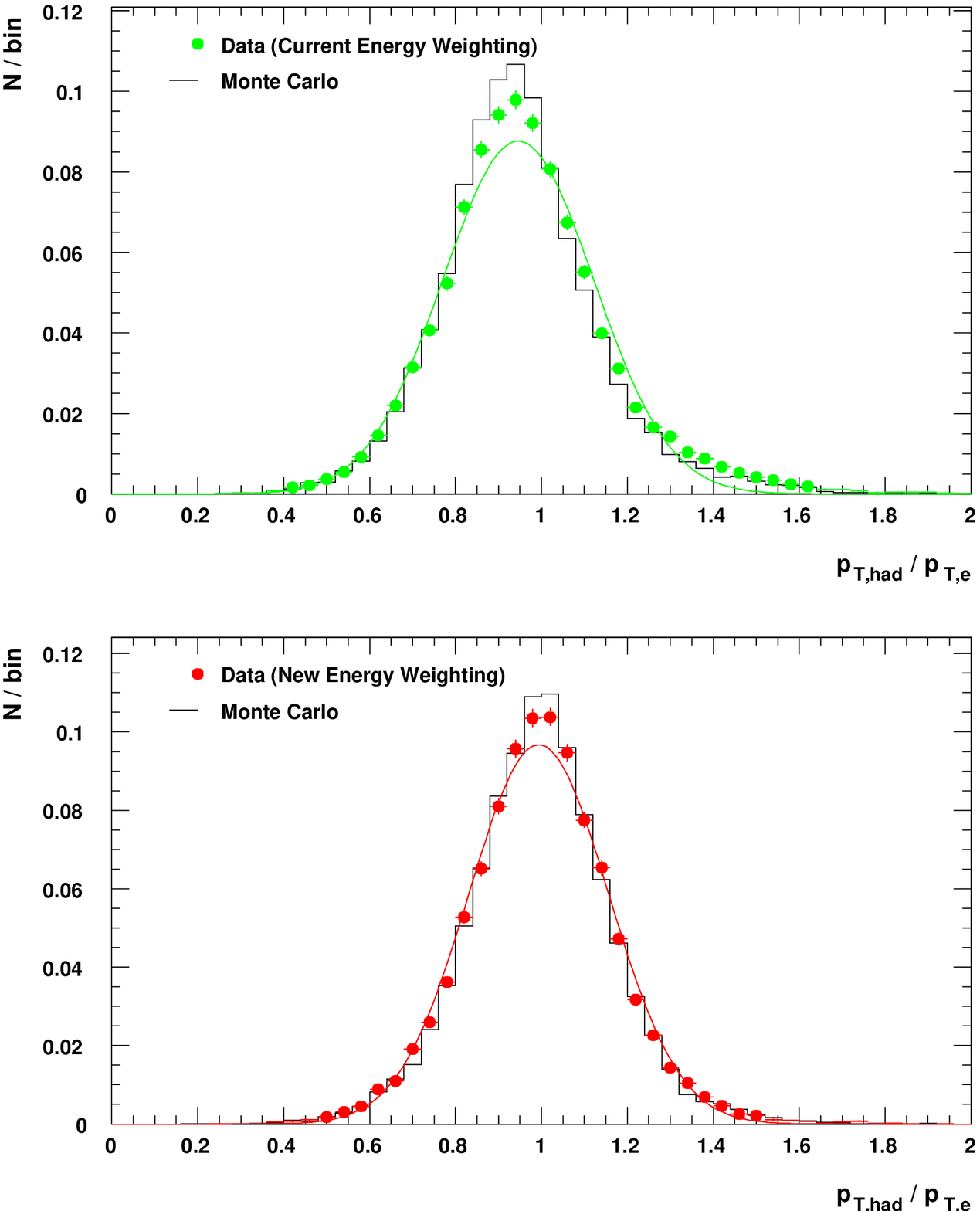}}
     \put(170,98){\epsfxsize=2.55in\epsfbox{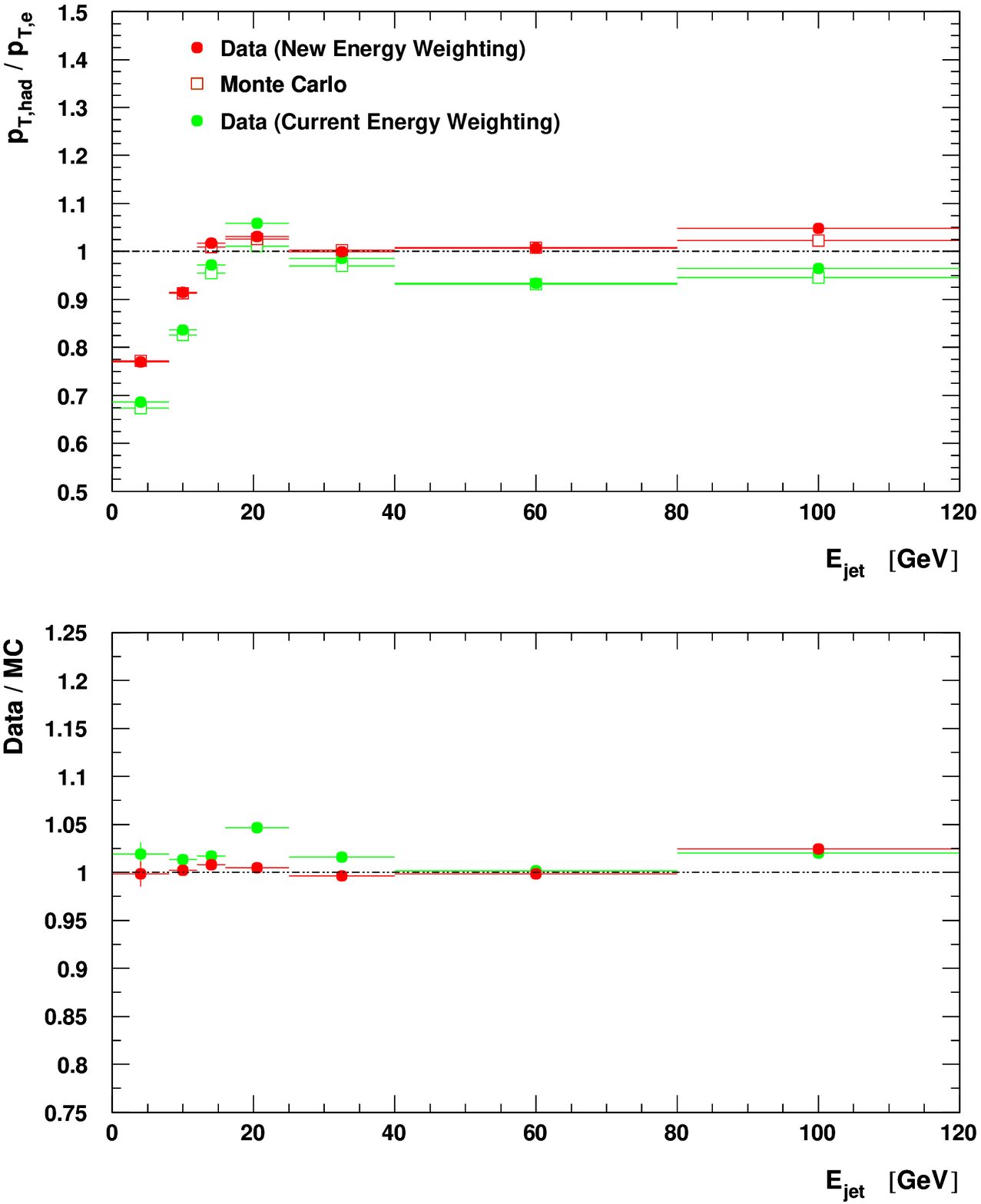}}
     \put(130,307){(a)}
     \put(130,190){(b)}
     \put(328,307){(c)}
     \put(328,190){(d)}
   \end{picture}
\vspace{-3.5cm}
\caption{Distribution of $p_t$-balance for the current (a)
  and a new energy weighting (b). The data are compared to the Monte
  Carlo simulation prediction (histogram) and to a Gaussian fit
  (curve). (c) Mean $p_t$-balance ratio as a function of the jet energy. Data (dark 
  points) and Monte Carlo simulation prediction (dark squares) in the
  new energy weighting scheme are compared to data (light points) and
  Monte Carlo simulation prediction (light squares) in the
  current energy weighting scheme. (d) The ratio of data and
  Monte Carlo prediction in the new (dark points) and the current
  (light points) energy weighting scheme as a function of the jet energy.\label{fig:weight}}
\vspace{-.2cm}
\end{figure}

Fig.~\ref{fig:weight} depicts the change in the $p_t$-balance
distribution if either the current (a) or the new (b) energy weighting scheme is
applied. In the current energy weighting scheme there are too many entries for
large $p_t^{\rm had} / p_t^e$ values which originate from neutral pions. In the new
energy weighting scheme there are fewer entries in the tails of this
distribution. Furthermore, the shape is more Gaussian-like and the
resolution is somewhat improved. 

The jet energy dependence of the
$p_t$-balance is also given in Fig.~\ref{fig:weight}. In 
the new energy weighting scheme an improvement of the hadronic energy
response at small jet energies can be observed (c) and the
agreement between data and Monte Carlo simulation is very good (d).


\section{Conclusions}

It has been discussed how physics data from electron proton scattering can be
used for the jet calibration of the LAr Calorimeter. In analyses of
photoproduction and NC DIS with high momentum transfer $Q^2$, the
calibration uncertainties are of the order $2\%$.

A new software weighting scheme was discussed. First applications in DIS
with high $Q^2$ show an improvement in the hadronic energy response at low
jet energies. Using this energy weighting scheme, the distribution of the
$p_t$-balance between the hadronic system and the electron gets more
Gaussian-like compared to the current weighting.


\section*{Acknowledgments}
I would like to thank the H1 Calorimeter and the Energy Scale Working
Groups for many discussions
and the opportunity to present these
results at this very interesting conference.
My thanks to the organizers.


\end{document}